CHAPTER 1

# THE ADELE-TEMPO experience : an environment to support process modeling and enaction.

Noureddine Belkhatir, Jacky Estublier and Walcelio Melo

Laboratoire de Génie Informatique
BP 53X, 38041 Grenoble Cedex 9 France

**ABSTRACT**

Process-Centered Software Engineering Environments (PSEE) have recently attracted a large number of researchers. In such environments the software processes are explicitly described and interpreted by the PSEE, allowing software activities to be automated, assisted and enforced. [Lehman and Belady, 1985; Osterweil, 1987] claim that this capability is a central element in a software development environment for the improvement of software product quality and software developers productivity.

We have addressed these problems in the framework of the Adele project. The Adele kernel, initially a configuration management system [Estublier et al., 1984], has been extended with respect to 1) modeling and support of complex product models: the Object Manager (OM) 2) modeling and support of software processes: the Activity Manager (AM), and 3) modeling and support of software product evolution: the Configuration Manager (CM). For data and product modelling, an ER/OO model has been implemented including SEE specific features;

On top of Adele kernel, which is a commercial product, we developed a Process Manager research prototype, Tempo, [Belkhatir et al., 1993] an enactable formalism based on two major concepts: objects may have a different description (role) depending on the process in which they are used, and processes are synchronized and coordinated by explicit connections.

ADL-Tempo is organized around the concepts of **software product**, **Work Environment** and **software process**. The software processes is the set of actions executed in Work Environements and which result in software products. We show how the Adele components: Object Manager, Activity Manager, Configuration Manager and Process Manager contribute, at their respective level of abstraction, to the support of products, work environements and processes and how their synergy provides a framework which simplifies subtentially the building of a process centered SEE.

1. **Introduction**

Significant academic and industrial research is currently being carried out with a view increasing productivity and improving quality during development and maintenance of software systems. When developing or maintaining large scale software projects programming-in-the-large [DeRemer and Kron, 1976] and programming-in-the-many [Floyd et al., 1989] must be taken into account. In fact, *The Process-Centered Software Engineering Environment* (hereafter PSEE) has been proposed to tackle such problems. Research in PSEE domain is diverse and covers many aspects. Some of the issues remain problematical, and a few attempts have been made to put forward a global approach to these problems. Adele is a product that makes a contribution to this field. It is a kernel in which the software processes can be supported explicitly.

Referring to SEE architecture and the integration technique classification presented in [Wasserman, 1989], Adele development focuses on data and control integration by developing an active software engineering database and, more recently following industrial experiments, on process integration.

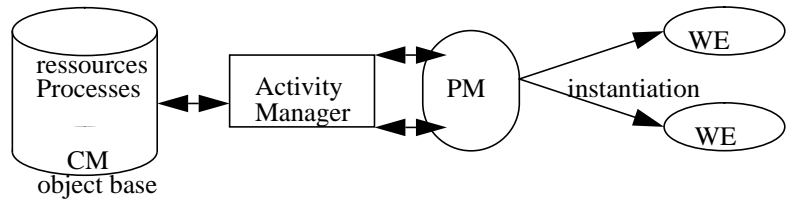

The figure shows the Adele-Tempo PSEE architecture. These are 4 main components:

(1)   An advanced object management system based on database technology (hereafter ADL-OM). In the ADL-OM simple and composite objects with attributes and relationships are explicitly described.
(2)   An Activity Manager (hereafter ADL_AM) where activities are modelled as occurrences of events which trigger actions according to ECA rules (Event-Condition-Action rules).

(3) A configuration manager (hereafter ADL-CM). Configuration management deals with activities related to software product evolution, i.e. their different versions, the selection of components and the actions that make them evolve i.e.

(4) Process Manager-PM (hereafter ADL-Tempo) as an academic research project which provides an enactable software process formalism. Owing to this formalism, specific software process models can be described and enforced. Each process instance performs activities in Work Environments(WE) whose goal is to produce Software products. For process support, a triggering mechanism and ECA rules have been implemented, including provisions for SEE's specific needs and for Work Environment a sub-database approach has been used. This formalism manages the communication and synchronization between teams and between members involved in the same project and controls the consistency of complex objects used simultaneously in different work environments by different software processes and agents.

Adele concepts and techniques have been influenced by object oriented Languages [Albano et al., 1993], database modelling [Hudson and King, 1989; Peckhan and Maryanski, 1988], trigger mechanisms [Dayal et al., 1988], and have led to progress in the following domains:

- v Type generic commands. Combining concepts from Entity-Relationship data models [Chen, 1976] and Object-Oriented technology [Atkinson et al., 1989];
- v Object type evolution [Zicari, 1989; Zdonik, 1990], management of structured and composite objects which evolve over time;
- v Relation types and their use for graphs, composite objects and constraint propagation [Boudier et al., 1988];
- v Extension of object orientation to capture contextual behavior [Dittrich et al., 1987];
- v Extension of ECA rules to express temporal triggers [Gehani et al., 1992].

We believe that a PSEE must manage the software items involved (Object Management), must provide support for process management (activity and process management). This article goes into the four aspects of Adele environment: ADL-OM, ADL_AM, ADL-CM and ADL-PM.

## 2. The Adele Object Manager

Various work on programming environments over the last ten years has made it clear that a database is the central component for managing software artifacts and an integral part of the Software Development Environment (SDE) and that standard database systems are not satisfactory [Belkhatir 87][Notkin 85] [Habermann and Notkins, 1986; Penedo, 1986; Unland and Schlageter, 1989]. SDE oriented database systems are not standard since database objects are not an abstraction from the real world but include programs, documents, and so on, and are managed by the database [Bernstein, 1987]

This characteristic removes inconsistencies that might appear between the data model and the real world since they are both equivalent. Any SEE managing either a model or files, is subject to failure [Meyer, 1985]. The database should manage:

(a)   large objects including many types.
(b)   highly structured and often compound objects (i.e. composed of other objects) with several versions possibly.
(c)   complex consistency constraints.
(d)   tools (compilers, link editors and so on) which may be applied to some objects and produce other objects.
      Furthermore, if the data base is to support processes, new requirements must be satisfied.
(e)   SEE is often concerned with activities which produce a product. Improvement of quality and predictability of software engineering work requires control and synchronization of multiple concurrent activities. There is an agreement that data integration is appropriate via an object management system.

Subsequently we defined the Adele kernel which covers all the characteristics mentioned above.

The Adele kernel is based on an entity-relationship database, complemented by *object-oriented concepts* [Atkinson et al., 1989] and an activity manager based on triggers.

### 2.1  The Adele data Model

The Adele Data Model[1] is derived from the Entity-Relationship model and integrates object oriented concepts. The information is organized as Objects and Relationships between them. An Object Type describes the structure and the behavior of an instance of the type. An Object is an instance of an Object Type in the OO sense; its identity is its (symbolic) name. A Relation Type describes the structure and behavior of a class of links between two objects. A relationship instance always has a Type and has an identity which is a triple: its Origin object name, its Relation type name and its Destination object name (O|R|D). A type (for

---

1. We present here the commercial Adele Data/Version model as distributed by Verilog Inc, 3010 LBJ freeway, suite 900 Dallas, Texas U.S.A. Verilog SA 52 av Aristide. Brian 92220 Bagneux France entities as well as for relations) can have one or more super types (i.e multiple inheritance) following the usual "inclusion set" semantics.

### 2.1.1 Attributes

An Object or a Relation Type contains a set of attribute definitions. An attribute definition is a name (identifier), an attribute domain (integer, date, boolean, string, enumeration) and other information such as the default value, the initial value, wether the attribute has a single or a set of values, wether its value is computed, substituted dynamically, etc.

Attribute instantiating occurs at any time. The type defines the allowed attributes, but does not impose attribute instantiating. Attribute definition follows the syntax presented below:

```
DEFATTRIBUTE
    suffix: (p, c, h, y, l) := c ;
    system : set_of (MSDOS, Unix*, VMS);
    machine COMP := "uname -n" ; -- Value is the result of the evaluation
```

Attributes are either built-in (name, date, `cmd,etc`) or defined by the user. Attribute scope is similar to variables scope in block-structured languages: the scope of an attribute instantiated to an object O is the object itself and all its `sub_objects`. This factorization of the information has proved to be very useful. Systems allowing the definition of complex objects should integrate this sort of mechanism.

### 2.1.2 Relations

Special attention is paid to relations management and control; they have indeed a significant role to play in a SEE, where objects are strongly inter-related. For example, for the relation "X program depends on Y interface"

- v A relation is an information (it informs us that X depends on Y).
- v A relation defines complex objects (X and its required interface can be seen as a complex object)
- v A relation defines a graph (here a dependency graph from X), hence a structure (tree, lattice, cyclic, acyclic)
- v If two objects are interrelated, an event on one may be propagate to the other. For example, if Y is modified, X may have to be re-compiled).

The following is an example defining a relation.

```
RELTYPE dep IS.... ;
  DOMAIN
    type = prog -> type = interface OR
    type = interface -> type = interface ;
  CARD N:N; DAG ;
  DEFATTRIBUTE ... ;
```

*Triggers*, *Methods* ....;
 END dep ;

Such an example means that the relation denoted *dep* is defined between a program and an interface, or between two interfaces. CARD expresses its cardinality: N mappings between objects of the relation are determined in both directions.Finaly, the *dep* relationship must be a Direct Acyclic Graph (DAG).

### 2.1.3 Aggregates.

An aggregate S is the object S and all the set of objects D related to S by a relationship R from S to D. In contrast to composite objects, components are sharable. Characteristics of S are not the characteristics of D, naming of components is independent, the existence of D is not necessarily related to the existence of S and the relation R must be defined by the user.

It is the user's responsibility to define how the nature of aggregates by determining the R characteristics. For instance, the existential, referential and composition integrity constraints [Boudier et al., 1988], the exclusivity of components, the method and constraint inheritance can be explicitly required when defining R. There is not really an aggregate concept but a wide spectrum from very loose aggregates (just objects in relation, not really an aggregate) to very strong aggregate.

### 2.1.4 Versions

The Adele Version Model is based on the branch concept. A branch models the evolution of a simple object and its attributes. It is a sequence of revisions where each revision contains a snapshot of the object attributes.

Different kinds of derivation graphs can be defined, establishing explicit relationships between branches; version groups can thus be easily defined. The Adele kernel provides the generic object concept (a branch is an object) and a mechanism for shared attributes. Arbitrary versions and composite objects are created and managed using explicit relationships between the different components.

However, version groups, versioned objects and composite objects are not explicit in the model. Their semantics and building are left to the application, even if it can be formally described by the Adele Process Modelling language.

### 2.1.5 Structurating and customizing: partitions and sub-projects

The Adele database is rather general, it is suitable for various projects, provided an adaptation is made for each of them. For large projects, sub-projects are highly independent and therefore may have different characteristics. For example, consistency constraints, languages, or tools can differ for both the kernel and some sub-systems; for configurations in maintenance and those under development, etc. A single data schema is not sufficient, even if most of the definitions are shared by most of sub-systems. In order to be able to define multiple data schemata, Adele introduces the notion of partition. This is defined as an aggregate built up from an object called a "root" and including the families and partitions connected to the root by the "part" relation ("part" is a composition relation). All components of a partition share the same schema. The "part" relation is defined "CARD 1:N; DAG;"; that is, partitions define a tree. Each partition inherits partition definitions (from its ancestor). The project is the root of the partitions tree, and is itself a partition. Each partition may modify (refine or enlarge) inherited definitions.

A sub-project is a partition defining an independent name space for objects. A partition plays a double role; it allows the definition of what is common between a set of objects (its common schema) and provides an abstraction mechanism, since a partition is an aggregate.

## 3. The Configuration manager.

Software Configuration Management is the discipline of controlling the evolution of complex systems[Tichy88]. On one hand, a Software Configuration Management (SCM) system must provide a powerful *object manager* system (OMS) allowing the software artifacts and their versions to be modeled and managed, on the other hand, a work Space manager for the specification and control of the activities such as designing, developing, documenting, building, validating and updating configurations and their components. An SCM results from harmonious collaboration between a Repository Space (RS) supported by the object manager and Work Spaces (WS). The WS controls the dynamic aspect of the SCM, i.e. the place where activities are performed; the RS is the place where software objects are modelled, stored and manipulated.

There are clear overlaps between PM and CM. However, we have limited CM to the following aspects at least in so form as this article is concerned :

v   Building configurations i.e. finding the list of configuration components.

v   Work Space management, which, from a CM perspective, covers two aspects:

> v   A WS must be (usually) a mono-versioned space; into which versions must fit. The SCM should be the only components involved in version control.
>
> v   A WS must contain the usual files and directories, while the OMS contains entities governed by a complex data model. How to relate objects between the OMS and the WS.

We shall rapidly run through these two aspects.

### 3.1   Building configuration

It is recognized that one of the major difficulties in SCM is to find the right list of components that constitute the required configuration. In Adele we designed a configuration builder, based on a specific product model, which is able to find out what the components of a complex configuration are. The principles of this configuration builder are presented below.

### 3.1.1 **The configuration Product Model.**

The Adele configuration builder relied on a predefined product model. This model was designed in order to represent modular software products and automatically compute software configurations.

This model borrowed the `module` concept from programming languages: the association of an `interface` with a body or realization. The distinction between interface and realization is interesting for numerous reasons, and has been widely recognized [Parnas72, Shaw84, Kamel87]. This Product model has three basic object types: family, interface and realization. A `family` object maps the `module` concept. Families allow interfaces to be grouped together. It gathers all the objects and all their variants and revisions related to a module. A family contains a set of interfaces.

Interfaces associated with a family define the features (procedures, types., etc.) exported by that family. An interface contains realizations [Haberman86]. Realizations are functionally equivalent but differ by some other (non-functional) characteristics. None is better than the others, none is the main variant, each one is of interest and needs to be maintained in parallel.

An is_realized relationship is defined between an interface and each one of its realizations. The dependency relation is also predefined, X and Y being variants, X depends_on Y is X needs Y directly (usually to be compiled). It often simply mimics the #include directive found in programs.

### 3.1.2 **Configuration Model**

Building a Configuration in Adele involves selecting a complete and consistent collection of objects: complete with respect to the structure of a Software System defined by the dependency graph closure and consistent with selection rules. Since the dependency relation is defined over variants, a System Model contains variants; it is a generic configuration [Estublier 88].

The depends_on relation is an AND relation (all destination objects must be selected) while the is_realized relationship is an OR relation (only one destination must be selected). The dependency relation can be an acyclic graph.

To be consistent, a System Model must satisfy (at least) the following requirements:

(1)     Only one interface per family is allowed.
(2)     Nodes must satisfy the constraints defined by the SM.
(3)     Nodes must satisfy the constraints set by all node ancestors in the graph.

Point 1 ensures that functionalities will not be provided twice. Point 2 ensures that all components are in compliance with the characteristics required by the model. In Adele, the characteristics required by the model are expressed in a first order logic language (and, or, not connective) which constrains component properties. Examples of such constraints can be "[recovery=Yes] and [system=unix] and [messages=English]".

Point 3 ensures that all components are able to work together. All compatibility constraints between components are associated with the component ancestor in the graph. Such constraints in Adele use the same language and can for example be "if [arguments=sorted] then [system=unix_4.3] or [recovery=no]".

### 3.1.3 Configuration instance.

An instance of a generic configuration, also called a bound configuration, is a set of revisions, one for each variant of the generic configuration. The instance is defined by the constraints to be applied in order to select the convenient revision for each variant. This selection is made in Adele on the basis of the revision properties, using the same first order logic language. An example of this sort of expression could be "([reserved=Riad] or [author=Riad] or [state=official]) and [date>18_02_89]".

In Adele a configuration is a standard realization: its interface is the root of the graph; its source text contains the constraints; its components are the destinations of the relation "composed_of"; its revisions are its instances. As a consequence a configuration also evolves as do variants and revisions in the usual way.

Evolution can be seen as a twin speed process. There is fast evolution, when a configuration process is used as baseline. This is everyday evolution; it covers bug fixes, minor enhancements and development, in other words all changes producing new revisions. Slow evolution is when a new configuration must be built. This is only necessary if new variants are to be incorporated in the model: new functionalities are needed, non-functional characteristics are desired or new internal constraints have been set.

The "current value" at time T of a configuration is its instance built at time T without constraint. Changes performed, in a configuration, by user Riad can be captured using attributes "reserved_by" or "author"; this is similar to DSEE [Leblang85, Leblang88] and ClearCase[Leblang93]. Any configuration can be built at any time by setting a cut-off date using "date < T" as a constraint and it is easy to reuse old revisions of the SM. A special configuration instance can be built using other revision attributes: "state=official and date<88_08_23" retrieves the official instance (if it exists) of this configuration as it was on August 23rd 1988.

### 3.1.4 Evaluation

This configuration builder and its associated product model were released in 1986, and have been used heavily by numerous customers. The experience gained revealed some drawbacks:

v  The default product model does not match all software easily. In particular, old products being designed with other structures in mind have problems in finding an interface/realization association.

v  The version definition and evolution constraints are often too rigid.

- Other constraints for building software configuration could be defined, using roughly the same process.
- The default product model does not easily support objects other than software programs.

These considerations led us to open up the data model, based on a general model for complex object management, a general version model including extended history, and a general process support.

### 3.1.5 Built Objects

Configurations are now generic. A configuration is a special case of a built object. The process, required to find instance components automatically, can be formally defined in any object type. This process is based on multiple graph closure, associative queries and the satisfaction of a set of consistency constraints. For example, a software configuration is built up from the closure of relations *depends_on* and *implemented_by* and by selecting realizations satisfying certain constraints. This work is under way but first experiments showed that the 3000 lines of code for computing configuration components, of the current configuration manager, are definable in one page of process definition. Only a few details cannot be formally described.

### 3.2 The CM Work Space manager.

In the previous section, we presented our data model intended to define the content of the **Repository Space (RS)**, i.e. where objects can be stored, and how they are versioned. It is the task of the Object Manager System to manage the repository space i.e. to represent, optimize, access, enforce protection and internal consistency of the RS. However, this is only the static aspect, storing and representing is not sufficient.

The dynamic aspect of configuration is performed in **Work Spaces (WS)**. A WS is the place where the usual software engineering tools are executed. The kind of objects that can be found in a WS depends on each system, but it must include files and directories, since they are the only entities known by the usual tools (editors, compilers, etc.). We can say the product model governing WS includes the File System (FS) model i.e. file, directory and link types.

Normally, a single version of files is present in a given WS, since standard tools do not recognize version concepts. It means that the RS, where all objects and their versions are stored, is different from the WS. The product model governing our RS allows us to model almost any concept, any kind of abstract and/or complex object, any version model, any software and team definition and history dimension can be defined. This contrasts with the simple model governing WSs.

It is clear for us that software engineering can only really be supported if WSs, RSs and Processes are closely controlled and managed, and *if they work in symbiosis*.

We will present the relationships between WS's and configurations, i.e., how activities in WS's and in the RS can be controlled and synchronized and the Adele propositions to solve the various problems caused by the duality of these two spaces.

### 3.3  Current approaches

This problem of maintaining a correspondence and consistency between both spaces (WS and RS) is a general problem. We now give a short characterization of the various solution classes, as used by commercial SCM systems.

v  The Check-in Check-out approach used by all systems based on the SSCS/RCS mechanism [Rochkind, 1974] [Tichy,85]. These tools are simple but have a very low level RS Model. No configuration or aggregate concept.

v  Explicit mapping by copies.In this class of tools, the mapping between WS and RS is explicitly managed by the CM, files are explicitly copied back and forth.The majority of CM tools belong to this class, but the WS model is very low (the File System Model) [Marzullo 88] and collaboration between WS is ill supported.

v  Sub-databases. A sub-database (sub-DB) is a sub-set of the central database; a change to an object is visible only in the current sub-DB, until "committed". Damokles [Dittrich89] and Orion [Kim91] are good examples. It allows cooperation protocols to be defined and concurrent activities to be supported. But a sub-database is not a WS, there is not WS support.

v  Virtual file system. Some tools sought to solve the mapping between WS and RS by "extending" a file system. This approach is followed by ClearCase [Leblang93], where the Unix FS driver is rewritten, and where any access to a file is interpreted by this new driver which decides the file version to provide. CaseWear [Cagan93] prefers to let users access the CM repository directly, but to solve the version problem, all tools need to be encapsulated.

v  Sub-database and Work Space. In this solution, the WS IS a sub-database. It means that all the DBMS services are available in each WS (local version, queries, short transactions, protection, etc.). A WS being a sub-DB, each individual user has the illusion of working alone on his objects. We have both the simplicity for end users and considerable modeling and control power.

Adele supports this last approach, but it is currently only partially implemented; our ambition is to provide a complete solution in the near future.

### 3.4  Work Space support and sub Data Bases

To satisfy the two opposite requirements: a WS model close to a bare File System (FS) and the RS with good modeling power, Adele supports the approach whereby a WS corresponds to a sub-database. A sub-database is a sub-set of the RS, formally described in the Product Model. A sub-database has also a (partial) File

System representation, which is a WS. The CM system is responsible for maintaining the relationship sub-DB <-> WS.

This relationships offers users and tools working in the WS, all DB services, such as a long transaction, recovery and so on, while still in a standard Unix environment.

In the following we present the WS <-> sub-DB (RS) relationships and how they are implemented using the *context* concept and specific mechanisms like file-mode, links and encapsulation strategies.

### 3.4.1  Context

A context is the set of entities contained in one or more aggregates. The aggregate is a very general concept, and almost all objects are aggregates: a sub-project, a document, a configuration, are examples of aggregates. Any aggregate (or set of aggregates) can be used at any moment as a context. When using a context, a user only "sees" the objects pertaining to that context.

A context can be checked in and out globally; it builds a WS, i.e. the sub tree of files and directories corresponding to the structure of the corresponding software product.

From the DB point of view, the context contains the aggregates; it is a sub-database; from a WS point of view a context contains a sub tree of files and directory, it is a WS.

A context is very closely related to the aggregates it is built up from; as soon as the contents of the aggregate evolve (new or deleted components), the content of the context evolves accordingly. Conversely, a context is loosely coupled with its WSs; adding/removing/changing components in a context propagates as changing/removing/changing the corresponding files in the WSs, bur an user request; in the same way, adding/removing/changing files in a WS is propagated, at user request by adding/removing/changing the corresponding components in the context (and hence in the corresponding aggregates).

The context is the concept which enables sub-DBs and WSs, to be built and to maintain the relationships between them.

This very general concept allows simply multiple policies to be implemented. For instance a WS can be built from configuration V2.3 (or rather context V2.3), with Change_sets C1 and C2 (other context). This is the way the change set approach is supported by Adele.

### 3.4.2  File mode and recovery

Once checked out in a work space, Adele records which version of each file is present in the WS, and almost any database command can be executed providing as parameter a WS file name. It is thus possible to ask for the attributes or relationship of a file, to create user defined relationships between files and so on.

Thanks to the context which maintains the correspondence between the WS and the RS, the WS model is extended by (a sub-set of) the RS model, without any need for most users to learn about the DBMS naming, version and product schema. Usual database queries can be answered from the WS perspective.

When a transaction involving changes in WSs aborts, all changes performed in the WS by that transaction are undone (roll-back). The WS is treated as part of the Database, and is covered by the usual DBMS features: transaction support, associative queries, interface, and so on. Protection, conversely, is only enforced for symbolic copies (see below).

### 3.4.3 Static and Dynamic links

In a WS, checked out files can be real file copies or symbolic links to the Adele database. Furthermore, links can be static or dynamic. A static link refers to a specific revision of an object, while a dynamic link always refers to the last available revision.

This feature allows only files requiring modification to be copied. ensuring protection of common parts (headers, libraries, etc.). Dynamic links enable small teams to rapidly synchronize and integrate their work as soon as (stable) changes are available.

### 3.4.4 Encapsulation strategies

Adele is not intrusive and no tools require encapsulation; a WS can be used as a bare Unix FS. By default Adele does not know what happens in a WS, there is absolutely no execution overhead, and absolutely no tool, method or habit change.

However certain users need to control the work performed in WSs more closely. Two approaches are proposed: 1) encapsulate tools and commands (all actions are interpreted by Adele and translated into corresponding basic actions), 2) Use the 3D tool to trap dynamically some file accessing. It can be used to perform "copies on write" strategy, tool activation monitoring, and better inter-WS support. Approach number 2 leaves the WS as a standard Unix space, without any encapsulation or change therefore, Conversely, it is sometimes difficult to infer which semantic actions are undertaken from only open/close commands.

Approach 1 has been in use for years by most of our customers; approach 2 is under implementation.

Using the different features presented above, a WS is a sub database, and a sub database is an RS.

### 3.5 WS management and Configuration

This integration of WS with RS and sub-databases allows almost any of the WS management strategies as presented in 2.3 to be implemented. The composition approach is our original approach, including selection of components, not only revision; the change set approach is implemented as presented above; the transaction approach is typically the sub database approach we have developed.

We have, in fact, developed different instantiations of Adele in order to implement different management strategies for different types of customer. These experiments and instantiations prove that Adele is really a generic configuration manager. Currently three standard instantiations are marketed, one based on the composition and transaction approach, another based on ADC philosophy, and a 3rd that mimics CMVC very closely

## 4. The Activity manager: Triggers support

### 4.1 Motivation

Software development activity and maintenance is a continuous process requiring different tasks. The activities to be managed are complex and various, for example design, documentation, constraints check, tool activation and policy control. Some activities are repetitive and well suited to automation [Bisiani et al., 1987; Conradi et al., 1991; Kaiser et al., 1988, Waters, 1989]. The Adele project has proposed an activity manager [Belkhatir 1987][Belkhatir et al., 1991] based on a trigger mechanism for managing these kind of activities. We describe the use of the Adele event-action mechanism. The example show how the Adele activity manager can specify and automate change management activity in a development environment.

### 4.2 Events and conditions

Event arise when a method is executed in the database, whether this method produces changes in the DB or not. Thus, methods calling other methods will dynamically give rise to more events.

In order to reduce the number of events created and to provide easier programming for users, the Adele concept of event is a mixture of events and conditions. It is a complex expression, involving the method which caused the events, temporal conditions, using the history mechanism, and external conditions

condition
```
EVENT delete_sensible = (!cmd = remove and (!object\comp/state =
released or !object@(status= validated)) ; PRIORITY 5;
```

This line expresses that event *delete_sensible* will be true whenever there is an attempt to delete a component (*!cmd = remove*), which is either a component of a released configuration *!object\comp/state = released* or which has been in the past in the status validated. The expression !*object* represent the name of the object receiving method !*cmd*. Similarly all parameters of the called method can be checked, as well as previous values of attributes and object when changed by the methods. It allows state transition machine and petri-nets to be easily programmed.

If multiple events are true simultaneously, their corresponding triggers are executed in the order of their priority.

### 4.3 Actions

An Action is a program in the Adele Language. An Adele language instruction can be a logical expression, an Adele command or a Unix command. This language is a simple imperative language, tailored to access Database information and navigate easily through arbitrary relationships. It is a meta substitution language (late binding of parameters and variables) that looks like the Unix shell, except that variables are multi-valued attributes, with provision for complex query and set operators.

A method can be called as part of the current transaction, and will be executed as a sub-transaction; it can also be executed as a separate transaction, synchronously or otherwise.

A method is declared in a relation type definition or an entity type definition. The implicitly associated event is the method call. Triggers and methods are inherited along the inheritance graph.

### 4.4 Coupling

A trigger program is executed each time the corresponding event is true. Four classes of trigger have been defined: pre-triggers executed before the method execution, post-triggers executed after the method. The set of pre-triggers, the methods and the list of post-trigger execution compose a transaction in the database sense. After-triggers and exception-triggers are executed after the transaction is committed or aborted respectively.

If compared with HiPAC coupling modes [Dayal88, 89], PRE triggers are *immediate* and POST triggers are *deferred*; separate transactions are *decoupled*. Adele adds AFTER and ERROR triggers which are decoupled, but executed after commit by the same process. In contrast with HiPAC, events/conditions and conditions/actions are evaluated with the same coupling, but the same event identifier can be present in different blocs, thus evaluated with different couplings.

### 4.5 Triggers and Data Model.

Following the same principle found in CLOS, Shood and other OO languages, triggers are inherited (they cannot be overloaded), and are executed from the most specific to the most general while methods are overloaded. This integration of event and trigger in the object type allows for better modelling of object behavior, since its triggers control are an important part of its behavior.

In contrast with most other approaches, e.g. Marvel [Barghouti and Kaiser, 1988], Epos [Conradi et al., 1991], where {pre, method, post} are declared all together, in Adele pre- and post- triggers are governed by events. This picture can therefore be more complex:

(a)     Pre- and post-triggers are not simple predicates, but can be arbitrary programs.
(b)     Pre- and post- are triggered by events, not necessarily a given command.
(c)     Different methods can share some pre- or post-conditions,

(d)     Events have priority, and triggers are executed in the priority order, by default the inheritance order is used, as in the previous example.

### 4.6  Active Relationships

Since our product model relies essentially on aggregates (configurations, subprojects, processes are aggregates), we found it fundamental to control aggregate behavior closely as well as aggregate interactions. To do so, triggers on relationships are able to control what happens both on their origin and destination objects.

Relationships are very similar to entities (they have attributes, triggers and methods); We have used active relationships to extend the Adele data model and include context-related behavior.

In a typical relationship, relation attributes are defined as the triggers and method to be acting when the relationship itself is a method parameter; for instance what to do when an instance is removed

        { ON remove rel DO {...}}

Adele relationships are binary, always defining an origin (named !O) and a destination (!D). In a typical relationship it is possible to define triggers and methods for executing on the origin and destination objects of a relationship.

For example, the composition relation can be defined in the following way:

```
TYPERELATION composition;
    1 ON ORIGIN delete DO {delete !D} ;
    2 ON DEST delete DO
        {print "you must delete first its container !O";
        ABORT};
    3 ON ORIGIN METHOD duplicate -d %new ;
        { copy !O -d %new };
    4 ON ORIGIN copy DO
        {makerel %new -r %realtype -d !D} ;
```

Line 1 stipulates that when deleting the origin of an aggregate (A), the destinations (X, Y, Z) must be deleted too. Line 2 stipulates that destinations cannot be deleted individually, since any attempt will produce an abort of the delete command. In line 3 we added a new method, duplicate, defined on the origin (A).

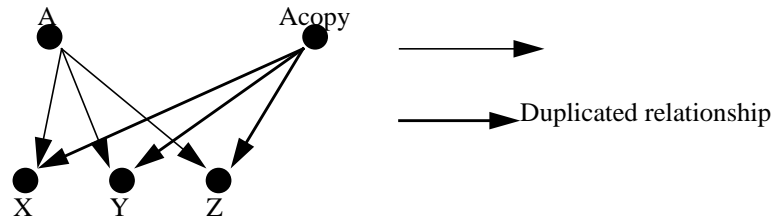

Thus, when "duplicate A -d Acopy" sentence is called, the method duplicate, which is defined in the composition relationship, will be executed. If duplicate is also defined in the A object type, it is dynamically overloaded by the one defined in the composition relation; if another relation type defines the same method on the same object instance, it is an error. The duplicate method only duplicates the aggregate head, in line 4, after the head duplication, the relationship composition is duplicated; this instruction is executed for each instance of the composition relation type, as we have shown at the end of the previous figure.

The composition relation defines an aggregate with logic duplication (sharing the content). Any other aggregate semantics can be defined in the same way.

### 4.7 Local and global trigger

A relationship is visible only if both the origin and destination objects are visible; as a consequence, depending on the context, a relationship can be visible or not, and consequently, its associated triggers are executed or not.

Suppose the **comp** relation, relating a configuration to its components, is defined in the following way:

 modif_rel_conf = (!modified=true AND !type=conf AND state=released) ;

 TYPERELATION comp ;
    POST DEST modif_rel_conf DO
        {print "cannot modify a released conf"; ABORT}

But the relationship comp is not visible in our contexts; the trigger will not be executed! A trigger can be either local, i.e. executed only if the relationship is visible, or global, i.e. executed whether or not the relationship is visible. In our case we must declare:

GLOBAL DEST modif_rel_conf DO {program}.

By default a trigger is local. A global trigger is executed with administrator rights, since the current user probably has no rights on external objects; for instance

GLOBAL DEST modif_conf DO { !O%state = obsolete } ;

means that when a configuration component is modified, the state of the configuration must become obsolete, whether or not the user who performed the change is aware of the configuration existence and has rights on it. Data consistency is enforced.

### 4.8 An example of shared resources using triggers.

Using the classical Philosophers and Spaghetti synchronization problem.

A philosopher needs two forks to eat spaghetti. He shares one fork with his neighbor. A philosopher may think, eat or wait to eat.

```
OBJECT Philo is rset;
    DEFATTRIBUTE
1       STATE : (eat, think, hungry) := think ;
2       METHOD eat DO newstate (self, eat);
3       METHOD think DO newstate (self, think) ;
4       ERROR ON eat DO IF STATE != eat THEN newstate(self, hungry);
END Philo;

OBJECT Fork is rset;
    DEFATTRIBUTE
5       STATE : (free, occupied):= free;
6       METHOD get DO IF STATE = occupied THEN {
            Print " The %name fork is occupied"; ABORT }
            ELSE newstate (self, occupied) ;
7       METHOD release DO newstate (self, free)
END Fork ;

RELTYPE Use;
    DOMAIN type = Philo -> type = Fork ;  CARD N:N ;
    PRE  ORIGIN eat  DO get (!D);
8   POST  ORIGIN think DO release (!D);
    AFTER  DEST release  DO
9       IF  ~!S%STATE = hungry  THEN
            eat  (!S);
END  Use ;

METHOD newstate (state) DO { mda self -a STATE state) } ;
```

   In this example, three objects types *Philo*, *Fork* and *Use* are defined modeling respectively the philosophers, the forks and the *use* relation ( which models the philosopher's arms. The eat and think methods (lines 2 and 3) only say that the philosopher state becomes respectively eating or thinking. Error trigger (line 4) states if the *eat* method fails (whatever the reason), the philosopher state changes to *hungry*. The fork method *get* (line 6), stipulates that if the fork is already in use, the current transaction must fail (ABORT); this can be one of the reasons why philopher  gets *hungry* (line 4).

   Relationship Use contains most of the story. Before (PRE) a philosopher (the relationship origin : !O) begins to eat, we must get the fork (i.e. the destination: !D); then a philosopher wants to think, we release the forks (line 8). Finally, after a fork has been released, if the related philosopher is *hungry*, he should try again to eat (line 9).

   It should be noted that this solution works for any number of philosophers and forks. The process model is independent of  the number of  instances involved. Extensions like using glasses, sharing dishes, and so on can be performed on line, adding relationships with the corresponding semantics, without any side effect on the fork sharing.

### 4.9 Application of the activity manager for change management

Assume a software product can be managed by the Adele database and a set of workspaces (WS). A file can be copied (Checked-Out) in a work environment or referenced directly in the database by soft links. When a modified file is replaced (checked-in: new revision) it is immediately available in all WSs where it is tested. A revision becomes 'official' when validated in all WSs. In order to specify this, we define first the relevant events and their relative priority:

```
DEFEVENT
    Delete_Official = [!cmd = delete, state = official]  PRIORITY 1;
```

Priorities indicate in which order the events are to take into account (higher number first). A method is an implicit event, thus *validate*, *invalidate* etc. are also events.

```
TYPEOBJET prog;
    1   PRE ON Delete_Official DO ABORT ;
    2   AFTER ON validate DO Check_Official ;
```

This trigger which is associated with each program (type=prog) says what to do in case *Delete_Official* and *validate* events occur.

Line 1 means before ("PRE") executing the action which consists of removing an official object, it is aborted (ABORT): it is not possible to delete a program in an official state. The last line means that after a validate command, Adele has to check wether the revision can be set to the official state (user defined command check_official). A revision can be official only if the *status* of all the relations RefWS (referred as (self|RefWS|**)) is qualified as 'status=validated' (line 3).

```
METHOD Check_Official ;
    3   IF ~(self|RefWS|**)%status == validated
        THEN newstate (official) ;
END check_official ;
```

The triggers of a relation are processed when an event happens to the object destination of such a relation. The current object becomes the object source of the relation (the event has been propagated from the destination to the source of the relation). In Adele, a WS is represented by an object. For our application we consider the relations RefWS; that is, the source object is a WS. Destinations are objects the corresponding WSs refer to by a symbolic link (logical copies).

```
RELATION  RefWS;-- Propagation on relation RefWS
        POST
            ON replace DO mail .....
```

```
              ON validate DO
                  IF  !O%author = !curentuser
                  THEN  newstate (validated) ;
              ON invalidate DO
                  IF  !O%author = !curentuser
                  THEN  newstate (unvalidated) ;
         AFTER
              ON officialize DO mail .....
 END RefWS ;
```

After a replace operation, a message is sent to all the owners of a WS with relation RefWs on the replaced object. After validate or invalidate commands; the attribute 'status=validate' is set on the relationship between the WS that validate and the validated object. The whole "PRE" command "POST" is a transaction. Any failure undoes the command completely. In our example, every WS may reject a "replace" command while evaluating the pre-condition (PRE) or the post-condition (POST).

This simple example shows how it is possible to:

v   enlarge existing commands (*replace* in our example),

v    define new methods,

v   associate object type definitions with their consistency checks,

v   automate propagations.

## 5. A process formalism: Tempo

### 5.1 Introduction

Recent developments have shown the need for an integrated view of the large scale software development environment that takes account of the artifacts and the way they are produced [Osterweil, 1987]. To fulfill these requirements, the activities as well as the software objects in development and maintenance must be managed. We describe below the extensions that we are making to the Adele system which permit it to support high-level software process description and enacting.

We shall discuss an original approach to process modeling known as the object driven process model. Our attention is focussed on how Object-Oriented concepts, the role concept and event-condition-action (ECA) rules make it possible to represent process model. In the following, we first evaluate our early experience in controlling software process, secondly we present TEMPO, our new software process formalism; and finally compare our approach with other current work and evaluate the research.

### 5.2 Trigger evaluation

We have made considerable use of event-condition-action formalism for modeling software processes, and the Adele trigger mechanism to control their execution.

Although this approach has produced good results, it does feature a number of weaknesses [Belkhatir et al., 1991], including:

(1)  the formalism is difficult to understand for human designers;
(2)  trigger execution is not very easy to control;
(3)  triggers are distributed over multiple types for the same action.
(4)  no high level concept such as process, work space, connection, planning,...

Simple questions such as "where am I?" or "what is my next task?" cannot be answered.

In part, these problems are due to the concept level provided by the Adele kernel, i.e concepts such as work spaces, user coordination and synchronization policies are not known to the kernel. Secondly, the fragmentation of information into different object and relation types makes it difficult to understand all the consequences of an action. For human understanding, it would be preferable for it to be grouped together in one unit. The large number of possibilities: pre-, post-, after, local, and global triggers multiple inheritance and relation overload, provide a flexible system wherein object behavior can be defined precisely, but it may confuse users. A clear picture of what will happen during execution is not easy.

For these reasons and to overcome these drawbacks, we designed a formalism named, TEMPO, on top of the Adele system for the management of software activities [Belkhatir et al., 1993].

### 5.3  The Tempo process formalism.

Tempo is a software process programming language based on the *role* concept. A software process type is defined as a set of objects playing a role. Due to the fact that objects are potentially shared simultaneously by different software processes (in which they play different roles) the behavior of an object cannot be defined statically; it is context dependent, i.e the object behavior is dependent on where the object is manipulated. Faced with the problem of multiple behavior definition we have used a two layer approach: (1) a kernel providing a set of general purpose concepts and mechanisms, and (2) an enactable formalism for software process definition and control, oriented towards work space control, team coordination and synchronization.

TEMPO defines a process model based on the two concepts **role** and **connection**. A role [Belkhatir 1992] allows one to redefine dynamically the static and behavioral properties of objects when playing that role in a process; a connection expresses how processes collaborate. This language is presented in the following section.

A software process model is described as a combination of software process types. A **process type** identifies and describes a set of activities. A software **process instance** is carried out by one or more users in a Work Environment (WE), which is the tuple (WS, PM, Tools, User).

Its rules give the policies that govern the use of WE objects. Software processes are activities executing concurrently and asynchronously. In our case, the communication and synchronization protocol between software processes are described by event-condition-action rules, because our approach is event-based. Each manipulation made on an object during the software process generates events, and rules defined in a process type make it possible to take these events into account in their respective work environments. Thus, synchronization between work environments is accomplished by capturing and processing the events provoked by the manipulation of shared objects.

A software process type is modelled in the Adele-DB as a standard object type. Thus a process entity can be instantiated and connected with other processes or software entities. Owing to multiple inheritance mechanisms, a process type can be refined and specialized. New attributes, roles, methods and rules can be overloaded and modified. Therefore, process customizing will be achieved by process type specialization.

### 5.4 The role concept

In order to satisfy the software process requirements, we need to change the characteristics and behavior of the manipulated objects, i.e. depending on the process where an object is used, its role is different. We introduce the ***role*** concept to describe this kind of object customizing. A role type makes it possible to change the definition of attributes, methods, and constraints of an object type when used in a process type. That is, a role type is a new type definition (properties and behavior) of objects in a process type.

#### 5.4.1 Definition

A role is the set of object instances, having the same behavior and characteristics for a given work environment (process step instance). A role defines the common behavior and characteristics of its instances. Characteristics means valid attributes, while behavior means methods and constraints. There is no strict relationship between role and type: (1) an object instance plays a single role in a given process, (2) object instances of the same type may play different roles, (3) instances of different types may play the same role, provided their types are compatible. A role is defined by a name, a type, local attributes, methods and rules.

The example below shows how the module type (from the product model) is referenced by process types *WS-change* through the *view* role type, and by *WS-valid* process type through the *to_valid* role type. During *WS-change* process instance enactment, the *view* role will be bound to the *module* instances manipulated by this process instance. This example also shows how attributes can be added to satisfy the needs of a specific software process instance. The attribute status is defined to satisfy the requirements of the WS-change process, and the updates made inside the execution context of this step to status attributed will be limited to view role. In this way other software processes that have manipulated the same module instances will be not affected by these updates.

### 5.4.2 Modeling contextual behavior: adjusting methods

Each role can redefine the original methods or define new ones to adapt the behavior of the original objects to a work environment context. For example, the module type has methods which are independent of the context wherein the module instances are used. However, when a module instance is manipulated by a work environment, other methods may be needed, e.g., the compile method of the module type may be different in the implementation work environment (compilation flags, etc.). The behavior of the module instances manipulated by a software process step can be tuned in order to satisfy the requirements of this work environment. The figure shows the role view with two methods, compile and crossref. The compile method overloads the original one defined in the software product model, and the method crossref is added to that role. In this way, we describe the contextual behavior of objects.

The following example shows how the module type is customized inside a **development** WE by the **to_consult** and **to_change** roles, and in a **validation** WE by the **component** role.

```
OBJECT Module ;
    ATTRIBUTE
        state = tested, untested, available ;
    METHOD
        compile ...;      -- with -C option
END Module;

PROCESS development ;
    ROLE testing = unitary_testing ;
    ROLE to_consult = module ;
    ROLE to_change =  to_consult/(responsible=!username);
        ATTRIBUTE state = compiled,edited, ready;
        METHOD
            compile ... ;    -- with -g option
        AFTER ON compile DO test ....
END development;

PROCESS validation ;
    ROLE component = module ;
    ATTRIBUTE
        test_suite = test1, test2;
    ...
END validation;
```

Role **to_change** will be bound to those modules the current user is responsible for (current user name (!username) equals attribute "responsible"); whereas the role **to_consult** is bound to the other modules. That is, when an occurrence of the **development** process is created, all module instances will be first bound to role **to_consult**, then those modules with the attribute "responsible" will be moved to the **to_change** role.

### 5.5 Process instances and Work-Environments

We define a Work Environment (WE) as a process occurrence. It is a tuple

WE = (WS, PM, Tools, User)

with WS a Work Space i.e. the set of object instances the process will perform on, PM the process model, Tools the tools that will be executed on the WS objects, and User the user(s) allowed to work in this Work Environment.

A WS has the properties presented before: by default any change performed in a WS is local to that WS (isolation property of transactions). The PM describes only what the process does in the WS. This property makes it easier to define a PM.

In our example, the state attribute is extended and may now contain two additional values *compiled* and *tested*. A modification of the *state* attribute is local to the WS. In a *validation* WE, the *component* role is also bound to *modules*. Each role has methods which are used to adapt the behavior of the object to that WE. That is, a role can redefine the original methods or define new ones in order to customize the object behavior for the WE in which it is used. For example, the *module* type has methods independent of the WE where the *module* instances are used. However, when a *module* instance is used in a given WE, other methods may be needed, e.g., the method *compile* may be different in a *development* WE than in a *validation* WE.

A user may have different WE simultaneously, and different WE may be active simultaneously for different users.

### 5.6 Structuring software processes

In our formalism, a complex software processes step can be broken down in sub-processes until the desired level of detail is achieved. Thus a complex activity can be broken down into a hierarchy of less complex activities. However, no special semantic is provided to express this policy. The role facility makes it possible to define a sub-process as a standard (sub-)role. A role can access all characteristics of its sub-roles.

### 5.7 Process and Role connection

Each process defines what happens in a WE as if it were performing alone; which is clearly false. Our basic hypothesis is that numerous activities are carried out in parallel. Some of such activities are working towards the same goal (for instance a new release of a software product), some do not collaborate towards the same goal, but share objects which is, to some extent, collaborating in the evolution of these objects. In all these cases the relationship between the WE must be explicitly defined.

We do not support the current approaches where a software process is described only as a tree of embedded sub-processes; we claim that SEE must be seen as *a federation of collaborating WE*, each WE being an enacted process occurrence. It is our belief that the conceptual definition of the network of collaborating WE is the major weakness in current SEEs.

We assume that the role corresponds to the grain size with respect to collaboration. The collaboration role is defined by a relationship that express the semantics of the collaboration. We provide a library of the usual such semantic relationships: **notify**, (which sends a notification to the WE owner when an event **notify_when** happens), **resynch** (that re-synchronize two objects when event **resynch_when** happens, merge, duplicate, share, deadline, protect, and so on.

To illustrate how role collaboration can be defined, let us imagine the following scenario:

When a new release of a given software product must be developed, a general process, called **release** is created. An arbitrary number of development WE and a single **validation** WE can collaborate in this release process. Each **development** WE can change only certain objects and have read access to the other objects of the release.

The synchronization between development WEs is as follows: when a given module M receives the **ready** state in a **to_change** role, M copies in each other **to_change** role must be merged, and their owner notified. If M is in a **to_consult** role, the new M version automatically replaces the previous one, and notification is sent to the WE user.

A module M receives the **available** state only if all its copies have the **ready** state. When all modules have the **available** state, the **validation** WE can be created.

Worthwhile collaboration thus takes place between the development WEs and the **validation** WE, but for reasons
 space this is not developed into more deeply here.

The following example shows how this policy is described in TEMPO.

```
TYPEPROCESS release ;

1    EVENT ready = (state := ready) ;
     ROLE USER = PMmanager;
     ROLE implement = development ;
     ROLE valid = validation ;
     ROLE components = module ; {

     ON ready DO {
```

```
2    IF implement.to_change.%name.state == ready THEN
3        implement.to_change.%name.state := available ;
4    IF implement.to_change.state == available THEN new valid ;} } ;

5    TYPECONNECTION consult_change IS notify, resynch ;
6        CONNECT implement WITH implement
7        WHEN to_consult.name = to_change.name ;
8        EVENT notify_when  = ready ;
                 resynch_when = ready ;
     END ;

     TYPECONNECTION change_change IS notify, merge ;
         CONNECT implement WITH implement
         WHEN to_change.name = to_change.name ;
         EVENT notify_when = ready ;
                 merge_when   = ready ;
     END ;
```

**END release** ;

- ν Line 1 stipulates that when an object gets to the state ready, the event **ready** occurs.
- ν The line 2 sentence *implement.to_change.%name.state* evaluates the set of values of the attribute *state* of the object that produced the *ready* event, as found in all *to_change* roles of the current process. Operator "==" means set equality. Line 2 means that all copies of the object *%name* have the *ready* state.
- ν Similarly, line 3 says that all these object copies must take on the *available* state.
- ν The line4 expression *implement.to_change.state* returns all state values of all objects in all *to_change* role. Line 4 means that when all objects get the *available* state, a *valid* role (i.e. a validation WE) must be created.

A connection is a special kind of relationship assumed to be instantiated between pairs of role instances. A connection is intended to define how each pair of connected object is coordinated. It may be a data flow definition, a status consistency checking, notification, deadline control, message passing, object evolution control and so on. It must be emphasized that connections are not symmetric; for instance, a development WE may automatically want to get new versions of objects as produced in a validation WE, probably not the other way!.

Some of this basic behavior is provided in a standard way, as for example *notify*, *resynch* and *merge* in this case. Using standard inheritance mechanism, each connection can reuse these process fragments (line 5), and redefine, for instance, the events for which some behavior must be executed. Line 8 means that notification must happens when the object becomes *ready*.

The CONNECT clause expresses which pair of objects must be connected. Line 6 means that, for a given release process, two **implement** roles are connected by a **consult_change** connection. Only those instances satisfying the expression found after the WHEN clause are automatically connected. In line 7 instances of **to_consult** in the first **implement** role are connected with instances of the **to_change** in the second **implement** role having the same name in both roles: the shared instances.

Thus, depending on the connections, the activity performed inside a WE may or may not interfere with other activities carried out in parallel during the software process.

### 5.8 Evaluation of TEMPO

#### 5.8.1 The role concept

One may claim that this kind of contextual behavior can be achieved by standard Object-Oriented techniques. Roles and type look similar; this raises the question: can roles be implemented in terms of typing and sub-typing? Is the concept of role needed at all? We claim the role concept has the following properties:

v **Prevent type explosion**. A role, as well as a type, is a template applied to a set of instances sharing the same definition (static and behavioral). A given object instance can be simultaneously a member of different roles (classes). Both roles and types can be seen as a viewing mechanism since a given object instance has a different description depending on the role (class) from which it is managed. One would need to create a sub-type for all the possible combinations of roles for a single type (combination numbers can be very large!), and to change instance type dynamically each time a new role is applied to it. However, there is a fundamental difference:

> The association between an instance and its type is statically defined at instantiationsinstantiation time, while an instance can be dynamically bound to an arbitrary role at any time.

Furthermore, since the instance may be shared and play different roles simultaneously, dynamic typing cannot be used. We introduce the possibility to changing type dynamically. In an O.O. system the type definition is created first, and then the instances of the types. In TEMPO on the other hand, the instances are usually created first, and are dynamically associated, for a while, to a (set of) role.

v **Identity is not altered**. Since a given object instance can be simultaneously a member of different roles (classes) there are compatibility rules between the roles (types) allowed for shared objects. In our model, objects can change behavior depending on the context without changing identity.

v  **Schema evolution-version.** Schema evolution support is an important facility for an O.O. database. A system that needs to reorganize the database when its schema is updated cannot support evolution properly. The role concept naturally integrates a type evolution facility, since role types are similar to object types in O.O. languages. The role concept offers two kinds of evolution 1) the role definition can change, they are role versions 2) the objects can change role dynamicallyinstantiations.

### 5.8.2  The Event-Condition-Action Rules

Triggers, when used in the software product model, are also very useful for capturing integrity constraints. We felt that triggers, when attached to a role, are also an important feature for controlling the operations performed in the software processes. Thus, triggers, defined in the software product model, describe invariant constraints, and triggers defined in the process type (in the roles) define the policies to apply in a work environment of that type.

These rules control the work performed inside a work environment; they are executed in response to actions performed in the process.

There is broad agreement that process-centered SEE should provide explicit support for temporal constraints. Due to the long life duration of software processes we need management mechanisms for time constraint and traceability. As for supporting process enacting we need to be able to reason with respect to execution sequences. Temporal logic seems to be a natural way to express temporal constraints in a software development environment. From the experience gained in other domains such as real time applications, we claim that the use of temporal constraints is fundamental to control activities sequencing. We are investigating the introduction of temporal logic as a way to plan and schedule activities in software processes.

At this level we are concerned by the specification of the life cycle of software artifacts, activities'duration and timing constraints, coordination of activities enactment and evolution control.

The purpose is to extend triggers with time concepts based on modal logic that would incorporate the ability to interpret temporal rules; this will be based on the generalized history mechanism presented before.

The trigger mechanism provides the flexibility for defining complex constraints that may be verified as soon as the object base is updated or when the change mode during long transactions is committed. The trigger mechanism is used for constraint checking as well as for information propagation. Consequently, the trigger mechanism extended by temporal concepts makes it possible to reason in terms of long term events, and how the software processes are executed over long periods, temporal rules seem a natural way for modeling such kinds of problem and temporal triggers an appropriate mechanism for supporting and processing this kind of information.

### 5.9 Supporting the TEMPO model in the Adele system

This section describes the implementation of the TEMPO model using the Adele kernel.

We have used the Adele kernel as the virtual machine for interpreting and enacting our software process formalism. The Adele kernel is based on an entity relationship data model. The implementation of the TEMPO model involves the following mapping for translating an Object oriented description of the TEMPO model into an O.O. entity-relationship model with triggers.

- Process types are Adele object types
- Roles are implemented by active relationships
- Rules are implemented by relationship and entity triggers.
- Methods and ECA rules are directly translated into Adele language.

## 6. Conclusion

The TEMPO formalism is designed for describing software processes in particular to implement process coordination and resource sharing. Its main capabilities are:

- An Object Oriented model. The resources (i.e. roles) involved in a process are modeled as objects providing a set of operations and constraints. The resources allocation is controlled by a central workspace dedicated to version and configuration management.
- The concept of role, which provides a new object-customizing mechanism where an object can be involved in different situations with different behaviors. This approach is close to the delegation mechanism of O.O. languages.
- The concept of connection, which expresses explicitly the synchronization an coordination between processes.

TEMPO is implemented on top of the Adele system which is the resource manager. Software process modes described in TEMPO are translated into Adele concepts (typed objects and relationships, event rules). We make a considerable use of the activity manager to support the process enaction. In this way, methods can be associated with pre and post conditions to determine the order of execution of methods. Methods are executed only when the conditions are met.

## 7. Related work

### 7.1 PSEE database

Our experience is twofold: 1) using Adele in industrial environments and writing industrial processes; 2) developing very demanding prototypes like Tempo.

These experiences shown that PSEE needs highly specialized databases. A great deal of work has been invested in trying to produce a widely accepted platform. These efforts have so far proved fruitless.

Software engineers have designed databases for their own purposes. Logically, these databases can be seen as file system extensions to integrate concepts from the entity-relationship data model. PCTE [Boudier et al., 1988], CAIS [Oberndorf, 1988] and Adele [Belkhatir et al., 1991] are platforms of this type. The basic idea is that tool integration can only be effective when sharing a common data schema. This trend has led to a sharp boundary between the platform (a passive object server) and the tools and activities that must be built on top of it. No real services are provided for activity control or for the tool control.

On current trends, PSEE DB integrate both approaches: ERA DB extended with OO features with better support for versions [Zdonik, 1990], structuring and consistency control. Products like PCTE+ and Adele 2 are of that class. The PSEE approach usually starts from such a platform and adds process modeling, e.g. ALF [Derniame et al., 1992], TEMPO [Belkhatir et al., 1993].

However, we do think that process support needs extensions in at least the following directions:

- Cooperative work and Sub-databases. Many workers are striving towards a common goal and share objects whilst performing their tasks. It is clear that the data base should provide broad ranging support to the different cooperation strategies, instead of imposing one.

- Evolution and Flexibility. A very flexible type management (modifying type, multiple types versions, etc.) from the DB is a mandatory feature for supporting process evolution, and probably one which is very difficult to offer from a DBMS point of view.

- History, Traceability and Measure. It will become increasingly important to be able to measure and assess process models, in order to tune, customize and enhance process model. This aspect will require, that the underlying DBMS have the ability to provide pertinent information about the performance of current and past processes, i.e. a temporal support.

- Viewpoints and Contextual Behavior. Each agent has a specific view of the total process. Thus each entity should have different descriptions (static and dynamic) depending on viewpoints. Entities can be shared by different activities and consequently the data base must support multiple simultaneous description of entities.

- Enaction support. Active DBMS. It seems to be accepted that the data base must be active; it means the data base must react to some events. The data base should provide a full event manager closely integrated with the transaction mechanism and with a compositionally complete language i.e. a basic process engine.

Clearly, no DBMS currently support fully all these features; but we do expect Adele will support them in the near future.

## 7.2 Configuration Management

Software Configuration Management is the discipline of controlling the evolution of complex systems[Tichy88]. On the one hand, a Software Configuration Management (SCM) system must provide a powerful *object manager* (OM) allowing software artifacts and their versions to be modeled and managed and, on the other hand, a *process manager* (PM) permitting the management and modeling of the processes used to design, develop, document, build, validate and update configurations and their components. An SCM results from harmonious collaboration between a Repository Space (RS) supported by the object manager and Work Spaces (WS) supported by Process Management. The WS controls the dynamic aspect of the SCM, i.e. the place where activities are performed; the RS is the place where software objects are modelled, stored and manipulated.

An SCM must meet conflicting requirements: the needs for a powerful and dedicated OMS with high power modeling, while activities are performed in WSs providing as usual, files and directories. Until now all SCM tools have been deficient in one, at least, of these two areas. One extreme consists of SCM systems which offer good services for managing the Repository Space, i.e., a good OMS, but little control over the corresponding files at the other extreme are WS managers supporting as product model only files and directories (a file system model).

An SCM system should rely on three major components: an object manager controlling the Repository Space (RS) where software artifacts and their versions are stored, a Work Space manager controlling the content and activities performed in the Work Spaces (WSs), and finally a Process Manager controlling the tasks and activities performed in both the WSs and the RS.

It is claimed that SCM needs a powerful data model, a WS model, greatly extending a file system; explicit mapping between RS and WS including inter WS collaboration and an explicit Process model, controlling both the RS and the WS and their relationships.

Commercial SCMs first used the bare file system with no data model at all. Currently advanced SCMs use simple databases and offer some data and process modelling capabilities, but much progress remains to be made in both directions. Adele is specifically designed for the support of specific data models and full process support, but the ideal has not yet been attained.

## 7.3 Process management

Among the kinds of software process programming language that the software process community has used for process-oriented software engineering environments [Madhavji, 1991] the following may be mentioned:

- v  the rule-based modeling software process using precondition, activity and post-conditions, e.g. Marvel [Kaiser et al., 1990], Epos [Conradi et al., 1991];
- v  the procedural [Osterweil, 1987], models derived from programming languages, e.g. Appl/A [Sutton et al., 1990], Triad [Sarkar and Venugopal, 1991], Galois [Sugiyama and Horowitz, 1991]; and
- v  the behavioral approach centered on artifacts produced (activity productions) rather than the specific procedures used to produce these artifacts [Williams, 1988], e.g. HFSP[Suzuki and Katayama, 1991].

Although our approach is a combination of these paradigms, in our case, rules are derived from event-condition-action formalism and enacted by triggers. We describe software process as an aggregate of object roles (artifacts) and associate with each role pre and post-conditions to control the consistency of these object roles. From procedural approach, rule description could involve procedural functions and procedures. The basis of integration of these mechanisms is an object manager supporting inheritance, aggregation, late binding and identification of objects

## 8. Conclusion

We believe that the definitions of a twin level system has led to an innovative system for the programming of work environments. We have extended Adele-DB, in order to closely link the static aspect (persistent software objects management, versions, etc.) and the dynamic aspect which is software process mangement. One of the goals of that work is to provide the software team leader with a simple language for the definition of software process models.

The contribution of this paper is our description of the two levels:

- v  A resource manager with a set of basic mechanisms for the general problem of maintaining the consistency of objects used simultaneously, for different purpose, in different and distributed workspaces. This is the description of the resource manager based on an object oriented DB, extended by triggers. We have described the main Adele kernel concepts. Adele provides a model of objects for the construction, integration and control of software objects in multiple versions:
    - v  complex object representation, with versions and query facilities; activation support, including an activity manager based on event-trigger concepts;
    - v  various abstraction levels supported by partitions. An aggregate is seen globally or in more detail
    - v  expression of objects' operational semantics. User-defined commands may be defined and associated with object types.